\documentclass[runningheads]{llncs}
\usepackage[T1]{fontenc}
\usepackage{graphicx}
\usepackage{algorithm}
\usepackage{algpseudocode}
\usepackage[dvipsnames]{xcolor}
\usepackage{amsmath}
\usepackage{amssymb}
\usepackage{float}
\usepackage{multirow}
\usepackage{booktabs}
\usepackage{capt-of}

\begin{document}
\title{A New Deep Learning and XAI-Based Algorithm for Features Selection in Genomics}
\titlerunning{DL and XAI Algorithm for FS in Genomics}
%
\author{Carlo Adornetto\inst{1,2}\and
Gianluigi Greco\inst{1,3}}

\authorrunning{Adornetto C. and Greco G.}
%
\institute{Department of Mathematics and Computer Science, University of Calabria\and
\email{carlo.adornetto@unical.it}\and
\email{ggreco@mat.unical.it}}
\maketitle              
\begin{abstract}
In the field of functional genomics, the analysis of gene expression profiles through Machine and Deep Learning is increasingly providing meaningful insight into a number of diseases. The paper proposes a novel algorithm to perform Feature Selection on genomic-scale data, which exploits the reconstruction capabilities of autoencoders and an ad-hoc defined Explainable Artificial Intelligence-based score in order to select the most informative genes for diagnosis, prognosis, and precision medicine.
Results of the application on a Chronic Lymphocytic Leukemia dataset evidence the effectiveness of the algorithm, by identifying and suggesting a set of meaningful genes for further medical investigation.

\end{abstract}

\section{Introduction}
In the field \textit{functional genomics}, starting from the results of the Human Genome Project, the evolution of sequencing techniques provides big volumes of data for each single patient by taking advantage of the \textit{high-throughput} and \textit{next-generation sequencing} i.e., a set of time and cost-effective techniques for sequencing DNA and RNA. By means of them, it is possible to measure the expression of thousands of genes for each individual and hence to collect quantitative gene expression profiles (GEP) to be used for research and clinical purposes. But despite GEP datasets represent a valuable source of information in healthcare---they are indeed used for diagnosis, prevention, and precision medicine---their analysis results challenging for three main reasons. The first one is the \textit{course of dimensionality}: a genomics dataset typically consists of a very large number of features (genes) and a small number of samples (patients); the second problem concerns \textit{imbalanced classes}: in the analysis of different groups of patients, genomics data are often stratified in classes according to different pathologies. In most cases, there is a significant difference between the number of instances in each class; finally, sequencing data are typically collected from multiple sources, different laboratories, and sequencing tools. This results in \textit{noisy datasets} which are difficult to analyze~\cite{koumakis2020}.\\ 
In recent years, Machine Learning (ML) and Deep Learning (DL) have been widely adopted in this field, providing breakthrough results and meaningful insights into the relationship between genomics and cancer~\cite{alhenawi2022,bruno2020data}. 
Although still very promising, DL models are in general not immediately interpretable, meaning that it is difficult to understand the causal relationship between the inputs and their outcomes. This is an even more severe problem in the bioinformatics domain, where it is crucial to understand, for example, in the case of genomics, how the expression of a gene can affect the progression of oncological patients. \\
We propose a new algorithm, based on DL and Explainable Artificial Intelligence (XAI), for genomics whose aim is threefold: first, select the most meaningful genes for a regression/classification problem; second, provide a more accurate prediction model; third, quantify and evaluate the effect of features on the predictions, through XAI. We used our algorithm for the GEP analysis of Chronic Lymphocytic Leukemia (CLL) patients, identifying a meaningful subset of genes for the disease prognosis.
The following sections are organized as follows. First, we review the most relevant related works in Section~\ref{sec:related}, and we then give a formal definition of the algorithm in Section~\ref{sec:algo}. The application and the results obtained by the algorithm for the CLL study are discussed in Section~\ref{sec:cll}. Finally, directions for further research are proposed in Section~\ref{sec:conclusion}. \\

\section{Related Works}\label{sec:related}
A number of recent studies propose and evaluate new approaches for feature selection (FS) on GEP datasets for cancer diagnosis and prognosis\cite{alhenawi2022}. Such methodologies mainly aim at selecting the most informative genes, which are able to characterize classes and identify groups of patients. In this context, the adoption of XAI methods has started to gain momentum for interpretability purposes as well as to enhance FS\cite{graham2020,meena2022,karim2019}. A widely used approach to overcome the \textit{course of dimensionality} problem is to perform dimensionality reduction using AEs \cite{danaee2017}. While this has been proven to be effective, the encoding is typically a non-linear projection of the variables into a lower-dimensional space, which makes it difficult to provide the proper interpretations of the results. In this work we propose a novel approach, which uses AEs for selecting the most informative genes without any change into the original features space, hence enhancing the explainability of the results, and still exploiting the representation abilities of~AEs. \\
We moreover use an ad-hoc defined XAI-based score in order to iteratively select the features by taking advantage of the Shapely Additive ex-Planation method (SHAP)\cite{lundberg2017}, a cooperative game theory-based approach for computing the \textit{shapely values}. Such values measure, locally (at the sample level), the contribution of each feature to the predictions of an ML model. In particular, for a given sample $x$, the set of features $F$, the contribution of the feature $j\in F$ is defined as:
\begin{equation}\label{eq:shap}
	\phi_j = \sum_{S\subseteq F\setminus\{j\}} \frac{|S|!(|F|-|S|-1)!}{|F|!} \;[f_{S\cup \{j\}}(x_{S\cup \{j\}}) - f_S(x_S)]
\end{equation}
with $\phi_j\in \mathbb{R}$ and where $f_{S\cup \{j\}}$ and $x_{S\cup \{j\}}$ denote the prediction model and the sample considering the only subset of features $S$ without the $j$-th one. 
In words, SHAP computes the contribution of a feature by comparing the model predictions obtained with and without a feature, for all the possible combinations $S$. Since the computation of the Equation~\ref{eq:shap} is inefficient in the case of NN as a prediction model---a NN should be re-trained for each combination of features ($2^{|F|}$)---the authors demonstrate in \cite{lundberg2017} that shapely values can be computed by solving a weighted linear least square regression with the proper shapely kernel. Although we used such an alternative method, we omitted the details and focus on the only definition of shapely values.

\section{The Algorithm}\label{sec:algo}
\begin{figure}[t]
	\centering
	\includegraphics[width=\textwidth]{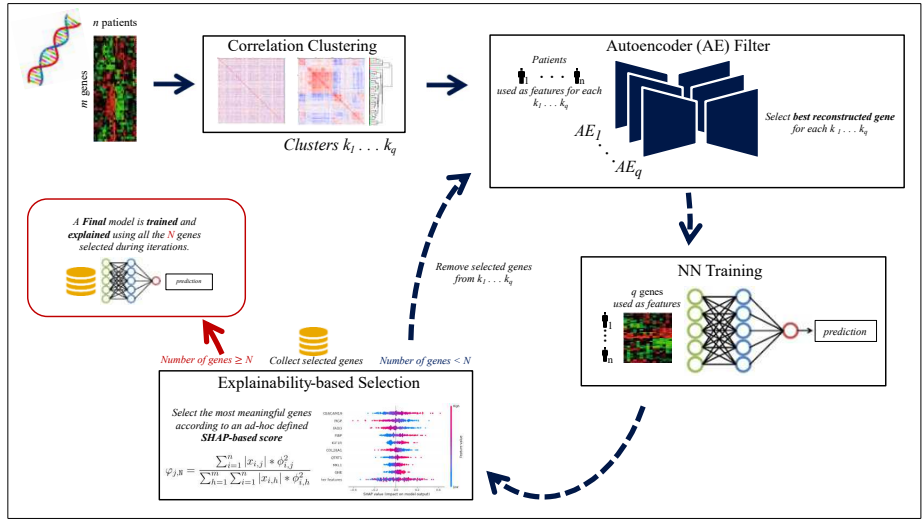}
	\vspace{-10pt}
	\caption{The Proposed Algorithm.} \label{fig:algo}
\end{figure}

The proposed algorithm is based on two main ideas: 
(1) we use a clustered correlation matrix in order to group features that enclose similar patterns and we then filter the redundant information for each group by using AEs. In contrast with previous works, in which AEs are used for dimensionality reduction, we still work at the level of the original features. In particular, we take advantage of the encoding and reconstruction abilities of the AEs assuming that the more accurate is the reconstruction of a feature, the more that feature is representative of the cluster it belongs to. We hence provide a more treatable dataset in terms of dimensionality, without loss of representativeness, by filtering redundant features; (2) we train NNs and we iteratively select the most meaningful features using a new ad-hoc defined SHAP score. We repeat the analysis by removing at each iteration, the previously selected features. We eventually use the set of selected features (from all the iterations) to train and explain a final model. Figure~\ref{fig:algo} shows the main algorithm phases.

\subsection{Formal Setting}\label{sub:formal}
Let be $\mathcal{D}=\{X,Y\}$ a dataset such that $X \in \mathbb{R}^{n\times m}$ is the matrix of inputs, and $Y\in \mathbb{R}^{n\times l}$  is the matrix of the corresponding labels. Let us further assume $m \gg n$ meaning that the dataset is characterized by a way larger set of features with respect to the number of samples.

As a novelty contribution, we introduce a new impact score, which, by means of the SHAP local explanation, measures the global impact of each feature on model predictions.
We hence associate to each feature (column) $j$ of $X$, used to train a model $\tt N$, a couple $(\rho_{j,\tt N}$ $, \varphi_{j,\tt N})$ were $\rho_{j,\tt N}$ is the correlation between the $j$-th columns of $X$ and their shapely values $\{\phi_{1,j},...,\phi_{n,j}\}$, and $\varphi_{j,\tt N}$ is defined as follows:
\begin{equation}\label{eq:intensity}
	\varphi_{j,\tt N}=\frac{\sum_{i=1}^{n}|x_{i,j}|*\phi_{i,j}^2}{\sum_{h=1}^{m}\sum_{i=1}^{n}|x_{i,h}|*\phi_{i,h}^2}
\end{equation}
With $\rho_{j,\tt N}$ and $\varphi_{j,\tt N}$ we want to emphasize \textit{how} and \textit{how much}, respectively, a feature globally affect the predictions of $\tt N$.

\subsection{Algorithm}
For sake of clarity, we introduce our algorithm by first defining a set of sub-procedures. The first one (Algorithm~\ref{alg:corr}) computes the pairwise correlation matrix $C \in \mathbb{R}^{m\times m}$ between the features (columns) of a generic real-valued matrix $M$. Finally it clusters $C$ in order to return a set $K=\{k_1,...,k_q\}$ such that for each $i=1,..,q$, $k_i$ is a set of indexes---a partition (cluster) for the columns of~$M$. \\
The second sub-procedure, defined in Algorithm~\ref{alg:ae}, trains an AE for each cluster, by using the transpose of the input matrix $M$---meaning that, for the AE model, each feature represents a sample and vice versa. The rationale here is that we assume the best-reconstructed feature (over the samples) to be the most representative of the cluster it belongs to. We denote $M_{k_i}\in \mathbb{R}^{n\times |k_i|}$ as a matrix including the only columns of $M$ which indexes are in $k_i$. The $evaluate$ function provides the column indexes of $M_k^T$ associated with the best-reconstructed feature. Finally, the sub-procedure returns a set $J$ of $q$ indexes---one for each~cluster.
\vspace{-35pt}
\begin{center}
\begin{minipage}{0.4955\textwidth}
\begin{algorithm}[H]
	\scriptsize
	\caption{Corr. \& Clustering}\label{alg:corr}
	\begin{algorithmic}
		\Function{CorrClustering}{$M$}		
		\State $C \gets corr(M)$
		\State $K \gets clustering(C)$ 
		\State \Return $K$	
		\EndFunction \\ \\ \\
		\vspace{1.5pt}
	\end{algorithmic}
\end{algorithm}	
\end{minipage}
\begin{minipage}{0.4955\textwidth}
\begin{algorithm}[H]
	\scriptsize
	\caption{AE Filtering}\label{alg:ae}
	\begin{algorithmic}
		\Function{AEFiltering}{$M$,$K$}
		\State $J \gets \emptyset$
		\For{$k \in K$}
		\State $\tt AE$ $\gets train(M_{k}^T)$ 
		\State $J \gets J \cup evaluate( \tt AE$ $, M_{k}^T)$ 
		\EndFor	\\	
		\Return $J$
		\EndFunction
	\end{algorithmic}
\end{algorithm}
\end{minipage}    
\end{center}
\vspace{-2pt}
The last sub-procedure, reported in Algorithm~\ref{alg:sel}, takes as input: the data, a matrix of shapely values $\Phi$ and the threshold $\beta\in \mathbb{R}$, with $\beta\in [0,1]$.
It first computes the correlations between each column of $M$ and the corresponding columns of $\Phi$. Subsequently, it computes the intensity for each feature following the definition of equation~\ref{eq:intensity}. It then selects the column indexes according to $\beta$ and the mean intensity, to finally provide a set $\tilde J$ of column indexes for $M$.
\vspace{-18pt}
\begin{algorithm}[H]
		\scriptsize	
		\caption{Selection}\label{alg:sel}
		\begin{algorithmic}
			\Function{select}{$\Phi,M,\beta$}
			\State $\pmb{c} \gets computeCorrelation(\Phi,M)$
			\State $\pmb{d} \gets computeIntensity(\Phi,M)$	
			\State $\mu \gets \frac{1}{|\pmb{d}|}\sum_{d\in\pmb{d}}d$
			\State $\tilde J \gets \{j \;|\; |\rho_{j}| > \beta \; \land \; \varphi_{j} > \mu, \;\forall\rho_{j} \in \pmb{c}, \forall \varphi_{j} \in \pmb{d}\}$ 
			\State \Return $\tilde J$
			\EndFunction
		\end{algorithmic}
	\end{algorithm}
 \vspace{-20pt}
\noindent The main procedure is described by Algorithm~\ref{alg:main}. After clustering the correlation matrix, it selects a set of meaningful features index to be added to $S$. It then removes the selected indexes from their corresponding clusters in $K$ and proceeds by repeating the analysis. Here we denote $X_J\in \mathbb{R}^{n\times |J|}$ (and accordingly $X_S$) as a matrix including the columns of $X$ which indexes are in $J$, and $\tt N$$_J$ (and accordingly $\tt N$$_S$) as a NN trained on $\{X_J,Y\}$. The iterative analysis stops when $\alpha\in \mathbb{N}$, $\alpha\leq m$ features are selected or on a maximum number of iterations. The algorithm eventually trains and explains a final NN using the set $S$.
\vspace{-16pt}
\begin{algorithm}
	\scriptsize	
	\caption{}\label{alg:main}
	\begin{algorithmic}
		\Require $X,Y$
		\State $K \gets$ \Call{CorrClustering}{$X$}
		\State $S \gets \emptyset$
		\While{$|S| < \alpha \lor \text{\textbf{not}}\, maxIter$}
			\State $J \gets $\Call{AEFiltering}{$X,K$}
			\State $X_J^b, Y^b \gets $\Call{dataBalancing}{$X_J,Y$}
			\State $\tt N$$_J\gets$ \Call{findModel}{$X_{J}^b,Y^b$} \Comment{Model Selection \& Training}
			\State $\Phi\gets$ \Call{Shap}{$\tt{N}$$_J,X_{J}$} \Comment{matrix of shapely values $\Phi\in\mathbb{R}^{n\times |J|}$}
			\State $\tilde J \gets$ \Call{select}{$\Phi,X_{J}$}
			\State $S \gets S \; \cup \; \{j_i \in J \;|\; i\in\tilde J\}$
			\State $K \gets K \setminus S$ \Comment{remove from their corresponding cluster}
		\EndWhile
		\State $\tt N$$_S\gets$ \Call{findModel}{$X_{S},Y$} 
		\State $\Phi\gets$ \Call{Shap}{$\tt{N}$$_S,X_{S}$}
		\State $J^* \gets$ \Call{select}{$\Phi,X_{S}$}
	\end{algorithmic}
\end{algorithm}

\section{A Use Case: Chronic Lymphocytic Leukemia}\label{sec:cll}
\subsection{Materials and Methods}
\begin{figure}[t]
	\begin{minipage}[H]{0.55\textwidth}
		\centering	
		\includegraphics[width=0.8\textwidth]{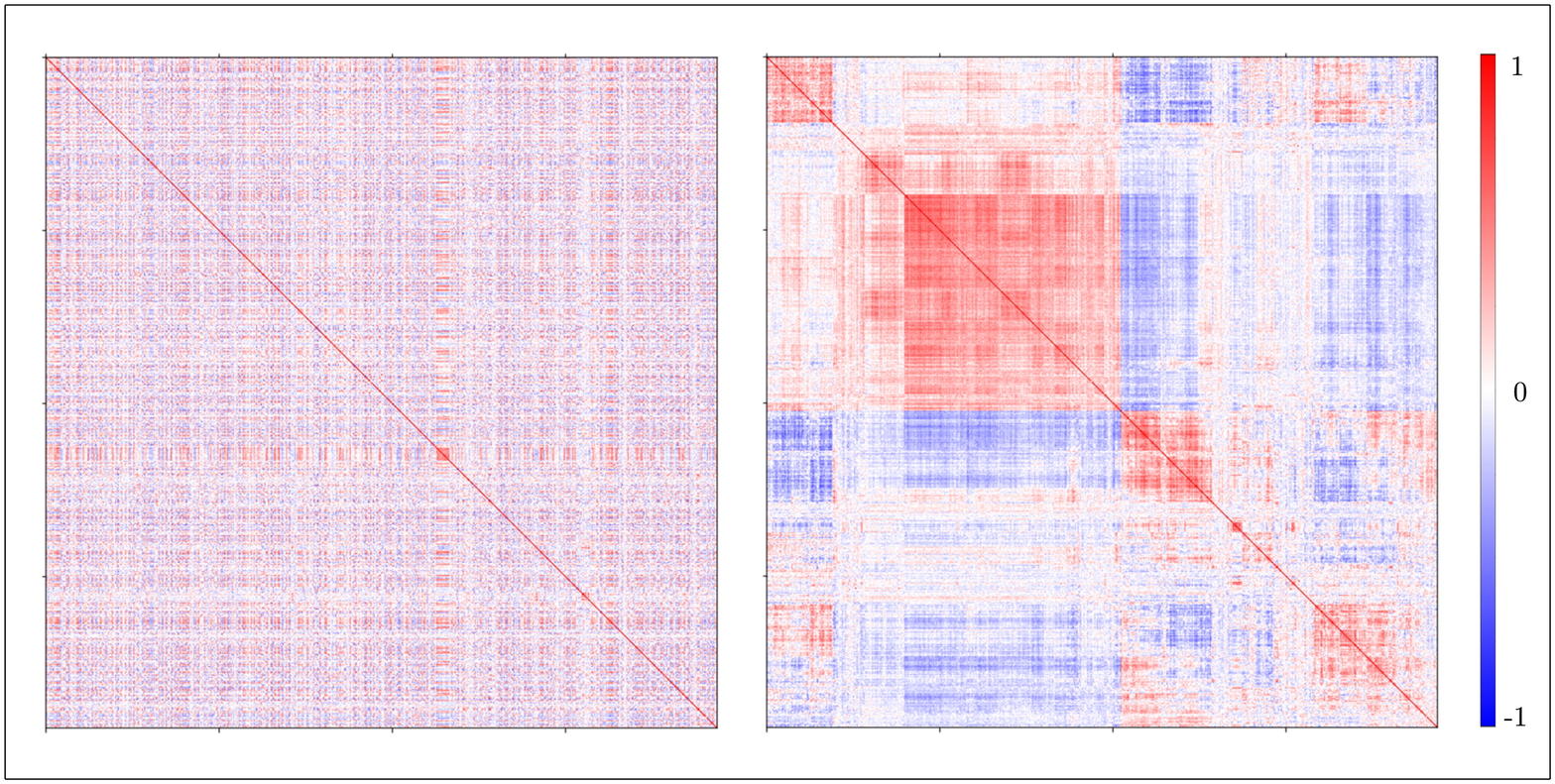}
		\vspace{-3pt}
		\caption{Genes clustered correlation matrix.} \label{fig:corr}
	\end{minipage}
	\begin{minipage}[H]{0.45\textwidth}
		\centering\scriptsize
		\begin{tabular}{|c|c|}
			\specialrule{0.7pt}{0em}{0em} \specialrule{0.7pt}{0.1em}{0em}
			\textbf{Iteration} &  \textbf{Accuracy (CI 95\%)}\\
			\specialrule{0.7pt}{0em}{0em} \specialrule{0.7pt}{0.1em}{0em}
			1 & 77.2\%-92.7\% \\ \hline 
			2 & 74.7\%-91.2\% \\ \hline
			3 & 68.6\%-89.3\% \\ \hline
			4 & 64.3\%-83.6\% \\ 
			\specialrule{0.7pt}{0em}{0em} \specialrule{0.7pt}{0.1em}{0em}
			final & \textbf{79.1\%-92.9\%} \\
			\specialrule{0.7pt}{0em}{0em} \specialrule{0.7pt}{0.1em}{0em}
		\end{tabular}
	\captionof{table}{Results over iterations.}\label{tab:acc}	
\end{minipage}

\end{figure}

We applied our algorithm for analyzing GEP of patients affected from CLL. 
The dataset used for this analysis is composed of 97 patients GEP for 19367 genes. 
The such dataset was extracted from the observational O-CLL1 study (clinicaltrials.gov identifier NCT00917540), where a set of newly diagnosed Binet A CLL cases were prospectively enrolled from several Italian institutions and studied for GEP~\cite{morabito2013,fabris2008}. 
For each patient a real-valued number is provided, indicating, as a factor of prognosis, the time interval after which the condition of the patient deteriorates. In particular, we distinguished two classes: \textit{class 0} (23 samples) for the patients whose condition deteriorates in a period shorter than 24 months, and \textit{class 1} (74) for the patients whose condition deteriorates in a period equal or longer than 24 months.
We used the proposed algorithm for training a NN to solve such a classification problem as well as to identify a set of meaningful genes over the whole set of 19367. We additionally provide insight into the prognostic power of such genes. The genes were initially clustered in 500 groups by using a hierarchical clustering technique---Figure~\ref{fig:corr}. The AE filtering selects 500 genes and we further applied a statistical filter in order to select 50 genes. After re-balancing the classes with the Synthetic Minority Over-sampling Technique (SMOTE), we perform model selection with 10-fold cross-validation in order to find the best (in terms of binary accuracy on the test set) NN for solving the classification problem. We finally use our SHAP scores (defined in Section~\ref{sub:formal}) to select the most meaningful genes, by setting $\beta=0.85$. After selecting a set of  $\alpha=50$ genes through the iterations of the algorithm, we use them to train and explain a final NN.
During cross-validation, the dataset was split into a train set of 138 patients (after re-balancing) and a test set of 10 only-real patients equally divided among the two classes.\\

\noindent The algorithm has been implemented using the Python (v3.8.11) programming language. NNs have been implemented by taking advantage of the Tensorflow (v2.6.0) framework along with the Keras library. XAI analysis was performed by means of the SHAP library \cite{lundberg2017}.

\subsection{Results}
The overall results are reported in Table~\ref{tab:acc}. In particular, for each iteration of the algorithm, we measured the accuracy of all the models obtained during cross-validation, for which we report the confidence interval. As we expected, the classification accuracy decreases with the algorithm iterations: the reason is that the previously chosen features---expected to be the most representative of each cluster---are no more considered for the subsequent analysis. An improvement in accuracy is instead reported for the final step of the algorithm, by which a model is trained using the set of genes selected during each iteration. The accuracy of the best final model is 100\%.

\begin{figure}[H]
	\centering
	\includegraphics[width=\textwidth]{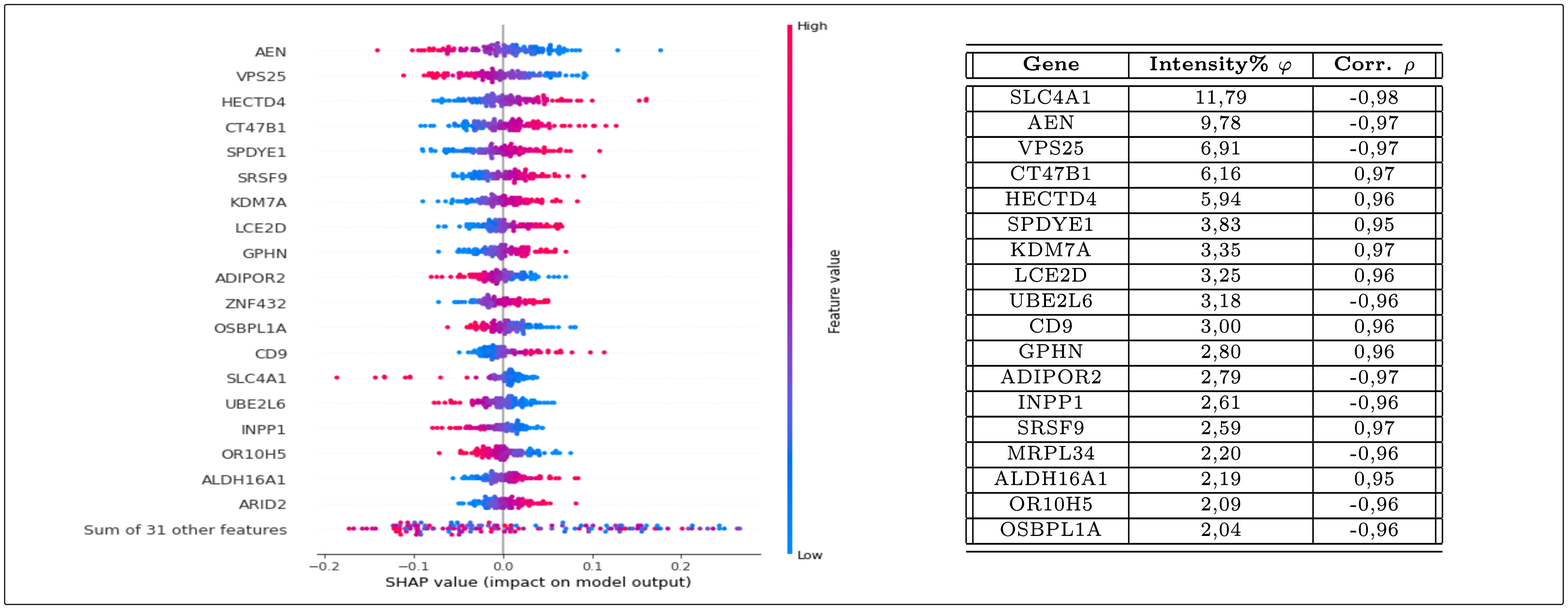}
	\vspace{-10pt}
	\caption{Final selected genes.} \label{fig:shap}
\end{figure}
\noindent Figure~\ref{fig:shap} reports, on the left side, a summarized representation of the shap values and, on the right side, the values for correlation and intensity for the most interesting genes found by our algorithm. 

\section{Conclusions}\label{sec:conclusion}
The algorithm proposed in this work can be used as a valuable tool in genomics to identify protective (or not) sets of genes for a disease, suggesting potential pathways for further medical investigation. A natural direction for future development is to perform a large-scale assessment of the algorithm performances, by using state-of-the-art benchmark GEP datasets.

\section*{Acknowledgments}\label{sec:akn}
We would like to thank Fortunato Morabito, MD (Biotechnology Research Unit, Aprigliano, Cosenza, Italy) and  Massimo Gentile, MD (Hematology Unit, Department of Onco-hematology, A.O. of Cosenza, Cosenza, Italy) and Antonino Neri, MD (Scientific Directorate, Azienda USL-IRCCS of Reggio Emilia, Reggio Emilia, Italy) for providing access to their clinical resources, which allowed us to gather valuable data for our analysis. Their support has been invaluable to us, and we are grateful for the time and effort they have dedicated to this project.

\newpage
\bibliographystyle{unsrt}
\bibliography{ref}
\end{document}